\documentclass[]{article}       
\usepackage{emulateapj}
\usepackage{multicol}
\usepackage{psfig}
\lefthead{PICKUP IONS AT THE TERMINATION SHOCK}
\righthead{ELLISON, JONES, \& BARING}
\submitted{The Astrophysical Journal, in press}
\def\today{\ifcase\month\or
  January\or February\or March\or April\or May\or June\or
  July\or August\or September\or October\or November\or
  December\fi
  \space\number\day, \number\year}
\def\teq#1{{$#1$}}
\def\dover#1#2{ {{#1} \over {#2}} }
             \font\sevenrm=cmr7

\def\asr{Adv. Space Res.}                       
\def\pss{Planet. Sp. Sci.}                      
\def\rgsp{Rev. Geophys. \& Sp. Phys.}           
\def\rpp{Rep. Prog. Phys.}                      
\def\ssr{Space Sci. Rev.}                       

\begin{document}

\newcommand{\vol}[2]{$\,$\rm #1\rm , #2.}                 
\newcommand{\figureout}[2]{ \figcaption[#1]{#2} }       

\title{DIRECT ACCELERATION OF PICKUP IONS AT THE SOLAR WIND \\
       TERMINATION SHOCK: THE PRODUCTION OF ANOMALOUS COSMIC RAYS}
\author{Donald C. Ellison}
\affil{Department of Physics, North Carolina State University,\\
       Box 8202, Raleigh NC 27695, U.S.A.\\
       {\it don\_ellison@ncsu.edu}}

\vskip 4pt

\author{Frank C. Jones \& Matthew G. Baring\altaffilmark{1}}
\affil{Laboratory for High Energy Astrophysics,\\
       NASA Goddard Space Flight Center, Greenbelt, MD 20771, U.S.A. \\
       {\it frank.c.jones@gsfc.nasa.gov, baring@lheavx.gsfc.nasa.gov}}
   \altaffiltext{1}{Universities Space Research Association}
\date{\today}

\begin{abstract}
We have modeled the injection and acceleration of pickup ions at the
solar wind termination shock and investigated the parameters needed to
produce the observed Anomalous Cosmic Ray (ACR) fluxes. A non-linear
Monte Carlo technique was employed, which in effect solves the Boltzmann
equation and is not restricted to near-isotropic particle distribution
functions.  This technique models the injection of thermal and pickup
ions, the acceleration of these ions, and the determination of the
shock structure under the influence of the accelerated ions.  The
essential effects of injection are treated in a mostly self-consistent
manner, including effects from shock obliquity, cross-field diffusion,
and pitch-angle scattering. Using recent determinations of pickup ion
densities, we are able to match the absolute flux of hydrogen in the
ACRs by assuming that pickup ion scattering mean free paths, at the
termination shock, are much less than an AU and that modestly strong
cross-field diffusion occurs.  Simultaneously, we match the flux {\it
ratios} He$^{+}$/H$^{+}$\ or O$^{+}$/H$^{+}$\ to within a factor $\sim
5$.  If the conditions of strong scattering apply, {\it no
pre-termination-shock injection phase is required} and the injection
and acceleration of pickup ions at the termination shock is totally
analogous to the injection and acceleration of ions at highly oblique
interplanetary shocks recently observed by the Ulysses spacecraft.  The
fact that ACR fluxes can be modeled with standard shock assumptions
suggests that the much-discussed ``injection problem" for highly
oblique shocks stems from incomplete (either mathematical or computer)
modeling of these shocks rather than from any actual difficulty shocks
may have in injecting and accelerating thermal or quasi-thermal
particles.
\end{abstract}


\section{INTRODUCTION}

It is believed that Anomalous Cosmic Rays (ACRs) originate as
interstellar pickup ions (Fisk, Kozlovsky, \& Ramaty 1974) which are
accelerated at the solar wind termination shock (Pesses, Jokipii, \&
Eichler 1981).  Such ions originate as neutrals that are swept into the
solar system from the external interstellar medium, and subsequently
ionized by the solar UV flux or by charge exchange with solar wind
ions.  Recent observations of pickup ions by the Ulysses spacecraft
(e.g.  Gloeckler et al.  1993) adds to the indirect evidence for this
scenario, which by now has become quite compelling.  However, one
essential element of the process, namely how pickup ions are first
injected into the acceleration mechanism, has engendered controversy.
We show here that standard and well-tested assumptions of diffusive
(also called first-order Fermi) shock acceleration allow the direct
injection and acceleration of pickup ions without a pre-injection
stage. We have employed our Monte Carlo simulation code (e.g.  Ellison,
Baring, \& Jones  1996) to study the physical parameters that the solar
wind termination shock must have in order to produce the observed ACR
fluxes.

\vskip 13pt

For input at the termination shock, we use a standard expression for
the shape of the isotropic pickup ion phase-space distribution based on
the derivation of Vasyliunas \& Siscoe  (1976) (e.g.  Gloeckler et al.
1993, 1994; le Roux, Potgieter, \& Ptuskin 1996), and normalize this to
the values reported by Cummings \& Stone  (1996) for the interstellar
ion flux in the heliosphere (see also Stone et al.  1996; Isenberg
1997).  We use the Cummings \& Stone fluxes, even though more recent
values have been reported (e.g.  Gloeckler 1996; Gloeckler, Fisk, \&
Geiss  1997), so we can make a direct comparison with their results.
In addition, since important parameters are uncertain at the
termination shock, we perform a limited parameter survey but always
find that we can easily match the observed flux of H$^{+}$, by varying
the strength of scattering.  For typical cases, we require that
$\lambda_{\parallel} \sim$ 5-10 $r_g$, where $\lambda_{\parallel}$ is
the scattering mean free path parallel to the mean magnetic field and
$r_g$ is the ion gyroradius.  This length scale of diffusion parallel
to the field seems fairly typical of that inferred in the vicinity of
planetary bow shocks (Ellison, M\"obius, \& Paschmann  1990),
interplanetary shocks (Baring et al.  1997), supernova shocks
(Achterberg, Blandford, \& Reynolds 1994), and that found in hybrid
simulations of quasi-parallel shocks (e.g.  Giacalone, et al.  1993)
but is much less than that found for the {\it undisturbed}
interplanetary medium (e.g.  Forman, Jokipii, \& Owens 1974; Palmer
1982; Moussas et al.  1992; Bieber et al.  1994; Gloeckler, Fisk, \&
Geiss 1997). If the turbulence we postulate for pickup ions is, in
fact, present, it implies that the termination shock generates fairly
strong, local magnetic field turbulence as has long been observed or
inferred at other collisionless shocks (e.g.  Lee 1982; 1983 and
references therein).  We are somewhat less successful in matching the
ACR flux {\it ratios}, He$^{+}$/H$^{+}$\ and O$^{+}$/H$^{+}$, seeing
less enhancement based on mass/charge than reported by Cummings \&
Stone  (1996).  We do, however, match the ratios to within a factor of
$\sim 5$, a relatively small difference given the uncertainties of
extrapolating flux densities to the termination shock and the
possibility that species-dependent heating or pre-acceleration could
occur in the solar wind before pickup ions reach the termination
shock.

Regardless of any uncertainty about flux ratios, that fact that we can
model the {\it absolute} hydrogen flux with no pre-acceleration is in
clear contradiction with the conclusions of most previous work
addressing pickup ion injection at the termination shock. For the most
part, previous work has argued that the highly oblique termination
shock would not be able to accelerate pickup ions directly. It was
postulated, for example, that some independent pre-acceleration phase,
perhaps at interplanetary shocks (e.g.  Jokipii \& Giacalone 1996) or
by second-order Fermi acceleration off Alfv\'en waves (e.g.  Isenberg
1986; Bogdan, Lee, \& Schneider 1991; Fichtner et al.  1996), or
transit-time damping of magnetosonic waves (e.g.  Fisk 1976; Schwadron,
Fisk, \& Gloeckler 1996), or shock `surfing' (e.g.  Lee et al.  1996;
Zank et al.  1996) was necessary before the pickup ions encountered the
termination shock and underwent their final acceleration to ACR
energies.  It has also been suggested that the termination shock was
not quasi-perpendicular for a substantial fraction of the time  (e.g.
Liewer, Rath, \& Goldstein 1995; Chalov \& Fahr 1996b) thus allowing
injection at times when the shock was less oblique.  Furthermore,
Chalov \& Fahr (1996b) and Le Roux, Potgieter, \& Ptuskin (1996)
suggest that reflected pickup ions from an already energized population
serve as seed particles for Fermi acceleration.  While it is certainly
possible that some pre-acceleration may occur or that the shock is
highly variable, our results indicate that the termination shock seems
fully capable of injecting and accelerating pickup ions directly in a
single step if standard diffusive shock acceleration assumptions are
made and if the self-generated turbulence is as strong as routinely
assumed in virtually all other astrophysical shocks which accelerate
particles.  Since diffusive shock acceleration predictions have been
tested extensively and successfully at directly observable shocks in
the inner heliosphere (and less directly at shocks outside the
heliosphere), we see no physical reason why the termination shock
should act differently, i.e.  should be incapable of generating
sufficient turbulence, or why standard shock assumptions shouldn't
apply (e.g.  Drury 1983; Jones \& Ellison  1991).  We note that claims
of extremely weak scattering of pickup ions ($\lambda_{\parallel} \sim
$AU) seem to be based on modeling of the {\it quiet} interplanetary
medium (Gloeckler, Fisk, \& Geiss 1997; Fisk et al.  1997; M\"obius et
al.  1998).  Convincing evidence that the mean free paths of pickup
ions with energies much less than ACR energies are $ \sim$ AU {\it
immediately behind} an interplanetary shock would, of course, make the
diffusive shock acceleration of these particles to ACR energies
impossible, since some particles must be able to diffuse back to the
shock from the downstream region in order to obtain MeV energies.

Our nonlinear shock acceleration model calculates the full distribution
functions of the various ion species at the shock including effects
from the shock smoothing produced by the back-reaction of accelerated
particles on the solar wind flow. The three most abundant ACR species,
H$^{+}$, He$^{+}$, and O$^{+}$, are included self-consistently in the
determination of the shock structure.  Since our Monte Carlo  technique
has not yet been generalized to spherical geometry, we are forced to
assume that the termination shock is plane.  However, the most
important process we investigate, the injection of pickup ions, occurs
locally and will not be seriously affected by this approximation.  In
addition, this implementation of the Monte Carlo simulation does not
treat solar modulation in a complete fashion, nor does it include
adiabatic losses; we anticipate including these in future work.  For
now, we artificially mimic the effects of adiabatic losses and truncate
acceleration by placing a free escape boundary upstream from the plane
shock.  We also neglect the presence of galactic cosmic rays on the
shock structure, relying on the estimate of Fisk (1996) that galactic
cosmic rays will not produce pressure gradients strong enough to smooth
the termination shock.

The most important model parameter we require is $\eta=
\lambda_{\parallel} / r_g$, the ratio of the particle mean free path
parallel to the magnetic field to the particle's gyroradius, and this
is chosen to match observed spectral intensities. For a given $\eta$,
our model gives the absolute normalization of all spectra at the shock.
Unfortunately, the absolute intensities are strongly dependent on the
pickup ion densities at the termination shock which are uncertain. Our
determination of the {\it relative} intensities of different ion
species, however, is not influenced in any important way by the
absolute normalization or by small changes in $\eta$, although changes
in Mach number can affect relative intensities as we show below.  In
partial support of the findings of Cummings \& Stone  (1996), we see
evidence for an acceleration efficiency that increases with $A/Q$ (mass
number to charge number). The actual values that we obtain, however,
are less than those inferred by Cummings \& Stone.  Similar $A/Q$
enhancement effects have been reported for diffuse ions observed at the
quasi-parallel Earth bow shock (Ellison, M\"obius, \& Paschmann
1990).

\section{MODEL}

The Monte Carlo  technique we use here has been described in Ellison,
Baring, \& Jones  (1996) and we refer the reader to that paper for
complete details.  Briefly, we have developed a technique for
calculating the structure of a plane, steady-state, collisionless shock
of arbitrary obliquity and arbitrary sonic and Alfv\'en Mach numbers
greater than one. We include the injection and acceleration of ions
directly from the background plasma and assume that, with the exception
of pickup ions, no {\it ad hoc} population of superthermal seed
particles is present.  The model assumes that the background plasma,
including accelerated particles, and magnetic fields are dynamically
important and their effects are included in determining the shock
structure.  The most important difference between the code we employ
here and that described in Ellison, Baring, \& Jones  is that we are no
longer restricted to subluminal shocks, i.e.  shock geometries where
the de Hoffmann-Teller (H-T) speed is less than the speed of light
(note, however, that our application in this paper to the termination
shock still focuses on subluminal shocks).  We no longer move particles
by transforming into the H-T frame, a frame where the {\bf u} $\times$
{\bf B} electric field disappears. Instead, we move particles in the
normal incidence frame and explicitly include effects of the {\bf u}
$\times$ {\bf B} field in translating the particles, allowing us to
model shocks of arbitrary obliquity. Apart from this generalization,
and the injection of pickup ions, the code used here is essentially
identical to that described in Ellison, Baring, \& Jones (1996).

The most basic assumption we make is that the complicated plasma
physics can be described by a simple scattering relation for individual
particles, i.e. ,
\begin{equation}
   \lambda_{\parallel}\; =\; \eta \, r_{\rm g} \quad {\rm or} \quad
   \kappa_{\parallel}\; =\; \dover{1}{3} \, \eta \, r_{\rm g} v \;\; ,
 \label{eq:mfp}
\end{equation}
where $v$ is the particle's speed in the local frame, $r_{\rm g}
=pc/(QeB)$ is the gyroradius of a particle of momentum $p$ and charge
$Qe$, $\lambda_{\parallel}$ is the mean free path parallel to the local
magnetic field, $\kappa_{\parallel}$ is the diffusion coefficient
parallel to the local magnetic field, and $\eta$ is a model parameter
which characterizes the strength of scattering and the importance of
cross-field diffusion. We assume $\eta$ is a constant independent of
particle energy, particle species, and position relative to the shock.
It's clear from equation~(\ref{eq:mfp}), that as a particle convects,
it will on average scatter after moving a distance
$\lambda_{\parallel}$ along the magnetic field.  We assume the
particles pitch-angle  scatter elastically and isotropically in the
local plasma frame regardless of their energy.  After each pitch-angle
scattering (which occurs every $\delta t \ll \lambda_{\parallel}/v$) a
new direction is obtained for a particle's velocity vector and a new
gyrocenter is calculated.  After some number of scatterings, a
particle's pitch angle will deviate by $\sim 90^\circ$ from it's
original direction and it will be gyrating around a field line  within
$2r_{\rm g}$ of the one the particle was circling originally.  Such
cross field diffusion is an integral part of diffusive acceleration at
oblique shocks (e.g.  Jokipii 1987; Ostrowski 1988) and Ellison,
Baring, \& Jones  (1995) showed that the scheme we employ here and in
Ellison, Baring, \& Jones  (1996) for cross-field diffusion, together
with the assumption contained in equation~(\ref{eq:mfp}), is equivalent
to a kinetic theory description of diffusion (e.g.  Axford 1965;
Forman, Jokipii, \& Owens 1974; Jones 1990), where the diffusion
coefficients perpendicular to ($\kappa_{\perp}$) and parallel to
($\kappa_{\parallel}$) the mean field direction are related via
$\kappa_{\perp} = \kappa_{\parallel} / (1 + \eta^2)$.  The parameter
$\eta$ in equation~(\ref{eq:mfp}) then clearly determines the
``strength" of the scattering and when $\eta \sim 1$, $\kappa_{\perp}
\sim \kappa_{\parallel}$ and  particles diffuse across the magnetic
field as quickly as they move along it (the so-called Bohm limit).  The
properties of highly oblique and quasi-parallel shocks tend to merge
when the scattering is strong (i.e.  \teq{\eta\ll 10}).

We simplify our model of the termination shock in several important
ways, namely we assume that the shock is plane and in a steady-state.
While the steady-state assumption is sensible for the termination shock
unless it is undergoing some form of perturbation on time scales short
compared to the acceleration time of the ACRs, it is less clear that a
plane shock assumption is valid for the curved termination shock.
However, the curvature of the termination shock will only be important
if the diffusion length of particles is comparable to the shock radius,
at which point high energy particles tend to leak away from the shock
and the acceleration ceases.  Otherwise the shock appears planar to the
accelerating particles.  Adiabatic losses in the expanding solar wind
are more of a concern, since these losses can be shown to set the
maximum energy particles obtain (Jones, in preparation).  For our work
here, we parameterize the maximum energy obtainable by placing a free
escape boundary at a distance upstream from the shock. The distance is
chosen to give maximum energies $\sim 100$ MeV, typical of ACRs.  We
find (we believe coincidentally) that the diffusion lengths of the
highest energy particles {\it along the magnetic field} are comparable
to the pole-to-equator distance.  Since our models generally have
\teq{\kappa_{\perp}\ll\kappa_{\parallel}}, the maximum scalelength
perpendicular to the termination shock (i.e.  in the radial direction)
is much shorter, of the order of a few AU.

Another important simplification is that we do not include a
cross-shock, charge separation potential in our model. A cross-shock
potential should exist and such a potential may have some effect on
injection. We leave this generalization to later work.

Once a satisfactory model for oblique shocks and shock acceleration is
developed, it becomes clear that the major problem with modeling a
given source is the array of parameters that are required.  Oblique
shocks are complicated and we do not see how they can be modeled
self-consistently without a large number of both environmental (e.g.
Mach numbers, shock speed, size, obliquity, etc.) and model (e.g.
$\eta$, type of scattering assumed, cross-shock potential, etc.)
parameters. While simplifications and a reduction in the number of
parameters can be made in some circumstances, the most extreme and
useful being the assumption of a plane-parallel shock ($\Theta_{\rm
Bn}=0$ everywhere, where $\Theta_{\rm Bn}$ is the angle between the
magnetic field and the shock normal), they cannot be made if the
obliquity and particle acceleration are important, i.e.  if the
accelerated particles are numerous enough so that a test-particle
solution is unrealistic. In this case, all of the parameters become
important and must be included.

\section{RESULTS}

The upstream parameters required for a solution are:  the shock speed,
$V_{\hbox{\sevenrm sk}}=V_{\hbox{\sevenrm sw}}$ (in the solar wind
frame), the magnetic field strength, $B$, the obliquity, $\Theta_{\rm
Bn}$, the temperature, $T$, the number densities of the various thermal
ion species, $n_{i}$, and the number densities of the various pickup
ions, $n_{\hbox{i}}^{\hbox{\sevenrm pu}}$, all of which are ambient
upstream conditions and can, in principle, be determined by
observations (we assume that the termination shock is stationary so
that the shock speed equals the solar wind speed, $V_{\hbox{\sevenrm
sw}}$).  Here,  ``upstream'' means far enough in front of the shock so
that backstreaming energetic particles do not influence the flow
parameters.  We also require our model parameter, $\eta$, and our
scattering assumption, equation~(\ref{eq:mfp}), which depend on the highly
complex plasma interactions that occur in the shock environs.  In
principle, these could be determined by comparison with observations of
space plasma shocks or 3-D plasma simulations; the prescription in
equation~(\ref{eq:mfp}) is a simple and transparent way to model these plasma
processes.

In addition to all of the above, we must also define the size of the
acceleration region by setting the distance (in units of mean free
paths), $d_{\hbox{\sevenrm FEB}}$, between the upstream free escape
boundary (FEB) and the shock. Accelerated particles that diffuse
upstream of the FEB are removed from the system, producing a high
energy turnover in the spectrum and giving a crude approximation of
adiabatic losses.  We emphasize that this complexity and array of
parameters is intrinsic to oblique shocks and must be included in any
realistic model.

\subsection{Parameters at the Termination Shock}

We use a simple model to relate values for solar wind parameters at the
termination shock to those at 1 AU.  Assuming that the solar wind speed
remains constant in its passage to the outer heliosphere, the density
of a solar wind ion species at the termination shock is
\begin{equation}
   n_{\hbox{\sevenrm i,TS}} \; =\;
   \left( \dover{1 AU}{D_{\hbox{\sevenrm TS}}} \right)^2
   n_{\hbox{\sevenrm i,AU}} \;\; ,
\end{equation}
the magnetic field at the termination shock is
\begin{equation}
   B_{\hbox{\sevenrm TS}} \; =\;
   \left( \dover{1 AU}{D_{\hbox{\sevenrm TS}}} \right)
   B_{\hbox{\sevenrm AU}} \;\; ,
\end{equation}
and the temperature of an ion species at the termination shock is
\begin{equation}
   T_{\hbox{\sevenrm i, TS}} \; =\;
   \left( \dover{1 AU}{D_{\hbox{\sevenrm TS}}} \right)^{2
        (\gamma_{\hbox{\sevenrm sw}} - 1)} T_{\hbox{\sevenrm i, AU}} \;\; ,
\end{equation}
where $D_{\hbox{\sevenrm TS}}$ is the distance to the termination
shock, the subscript ``AU'' indicates values at Earth, and the
subscript ``TS'' indicates values at the termination shock.  We have
assumed that the solar wind flux per solid angle is conserved, and that
the magnetic field strength decreases as $r^{-1}$ in the tightly-wound
Archimedean ``Parker'' spiral (Parker 1958); this field is dominated by
the tangential component, while the radial component drops off as
$1/r^2$.  Also, the temperature is determined by adiabatic expansion of
the wind, i.e.  $ {\cal V}^{\gamma_{\hbox{\sevenrm sw}} - 1} T = {\rm
constant}, $ where ${\cal V}$ is a volume element and
$\gamma_{\hbox{\sevenrm sw}}$ is the ratio of specific heats for the
solar wind.  We take $\gamma_{\hbox{\sevenrm sw}} = 5/3$ and assume
that the termination shock is at 85 AU in all that follows.  If, for
example, we take values at 1 AU of $n_{\hbox{\sevenrm p,AU}} = 8$
cm$^{-3}$, $B_{\hbox{\sevenrm AU}} = 5\times 10^{-5}$ G,
$T_{\hbox{\sevenrm p, AU}} = 2\times 10^{5}$ K, and $V_{\hbox{\sevenrm
sw}} = 500$ km s$^{-1}$, we have for the termination shock parameters:
$n_{\hbox{\sevenrm p,TS}} = 1.1\times 10^{-3}$ cm$^{-3}$,
$B_{\hbox{\sevenrm TS}} = 5.9\times 10^{-7}$ G, $T_{\hbox{\sevenrm p,
TS}} = 535$ K, and, for the Mach numbers, ${\cal M}_{\hbox{\sevenrm S}}
\simeq 130$ and ${\cal M}_{\hbox{\sevenrm A}} \simeq 13$ (${\cal
M}_{\hbox{\sevenrm S}}$ is the sonic Mach number and ${\cal
M}_{\hbox{\sevenrm A}}$ is the Alfv\'en Mach number).  For this
example, we have neglected pickup ions and ion species other than
protons. The addition of pickup ions will lower ${\cal
M}_{\hbox{\sevenrm S}}$ dramatically.  We assume here and elsewhere
that the electron and proton temperatures are equal, and that all ions
have the same temperature per nucleon.  This equality is used for its
expediency and can, of course, be relaxed if data shows otherwise.  In
reality the electron component of the solar wind is somewhat hotter
than the protons (e.g.  see Baring et al.  1997), perhaps due to their
greater conductivity; large scale averages for electron temperatures
are presented by Phillips et al. (1995).

We further assume a fixed value for $\Theta_{\rm Bn}$, repeating that
our's is a plane shock model and can only describe a shock with a
constant far upstream obliquity.  We note that, while
$d_{\hbox{\sevenrm FEB}}\ll D_{\hbox{\sevenrm TS}}$ in our models, the
size of our system along the field lines, $d_{\hbox{\sevenrm
FEB}}\tan\Theta_{\rm Bn}$, is comparable to $D_{\hbox{\sevenrm TS}}$
since $\Theta_{\rm Bn} \lesssim 90^\circ$.  We do not model the range
of magnetic field geometries around a spherical shock.

\subsection{Pickup Ion Contribution to the Sonic Mach Number}

Pickup ions contribute to the sonic Mach number ${\cal
M}_{\hbox{\sevenrm S}}$ through both their mass loading of the solar
wind, and also their velocity dispersion relative to the mean speed of
the wind.  This latter component is crucial to the determination of
${\cal M}_{\hbox{\sevenrm S}}$ at large distances from the sun, where
adiabatic cooling of the solar wind has diminished its pressure below
that of the pickup ions.  The sound speed in the solar wind frame is
\begin{equation}
   c_{\rm s} \; =\; \sqrt{\dover{\partial P}{\partial\rho}}
      \; =\; \sqrt{\dover{\gamma P}{\rho}}\quad ,\quad\hbox{for}\quad
      P\,\propto\,\rho^{\gamma}\,\propto\, {\cal V}^{1-\gamma}\;\; ,
\end{equation}
for a gas of one species (e.g.  protons or helium ions), where $\gamma$
is the ratio of specific heats for that species, $P$ is the pressure,
$\rho$ is the mass density, and ${\cal V}$ is the volume.  If the shock
speed in the solar wind frame is $V_{\hbox{\sevenrm sk}}$ ($\approx
V_{\hbox{\sevenrm sw}}$), then the sonic Mach number is
\begin{equation}
   {\cal M}_{\hbox{\sevenrm S}}\; =\;
   \dover{V_{\hbox{\sevenrm sk}}}{c_{\rm s}} \; =\; \sqrt{\dover{d}{\gamma}}\;
   \dover{V_{\hbox{\sevenrm sk}}}{\sqrt{\langle v^2\rangle}}\quad ,
\end{equation}
where the pressure, \teq{P=nm\langle v^2\rangle /d =\rho\langle
v^2\rangle/d}, is expressed in terms of the mean of the square (i.e.
dispersion) of the particle speeds $v$ (measured in the solar wind
frame).  Here \teq{d} is the dimensionality of the system.  For a {\it
phase space} speed distribution \teq{f(v)} of non-relativistic
particles,
\begin{equation}
   \langle v^2\rangle\; =\;\int_0^{\infty} v^4\, f(v)\, dv \Biggl/
   \int_0^{\infty} v^2\, f(v)\, dv\quad .
\end{equation}
For a monoenergetic pickup ion injection distribution \teq{f(v) = \delta
(v-V_{\hbox{\sevenrm sw}})}, we have $\langle v^2\rangle =
V_{\hbox{\sevenrm sw}}^2$ (\teq{=V_{\hbox{\sevenrm sk}}^2}), while for
thermal solar wind particles with \teq{\exp [-mv^2/(2kT)]}, the
familiar result \teq{\langle v^2\rangle =3kT/m} for a non-relativistic
Maxwellian emerges.

To accommodate the two-component population of solar wind and pickup
ions (denoted by subscripts ``sw'' and ``pu,'' respectively), the speed
of sound must by modified.  Since pressures and densities add linearly,
i.e.  \teq{P=P_{\hbox{\sevenrm sw}}+P_{\hbox{\sevenrm pu}}} and
\teq{\rho=\rho_{\hbox{\sevenrm sw}}+\rho_{\hbox{\sevenrm pu}}}, then
the adiabatic laws of
\begin{equation}
   P_{\hbox{\sevenrm sw}} {\cal V}^{\gamma_{\hbox{\sevenrm sw}}-1}
   \; =\; \hbox{const}_{\hbox{\sevenrm s}}\quad \hbox{and}\quad
   P_{\hbox{\sevenrm pu}} {\cal V}^{\gamma_{\hbox{\sevenrm pu}}-1}
   \; =\;\hbox{const}_{\hbox{\sevenrm p}}
\end{equation}
can be used to derive
\begin{eqnarray}
   c_{\rm s}^2 &\equiv & \dover{\partial P}{\partial\rho}
   \; =\;\dover{\partial P}{\partial {\cal V}}\;\dover{d{\cal V}}{d\rho}
   \nonumber\\[-5.5pt]
 && \label{eq:csoundsq}\\[-5.5pt]
   & = & -\biggl(\dover{\partial P_{\hbox{\sevenrm sw}}}{\partial 
   {\cal V}}+\dover{\partial P_{\hbox{\sevenrm pu}}}{\partial {\cal V}}\biggr)
   \;\dover{{\cal V}}{\rho}\; =\;\dover{\gamma_{\hbox{\sevenrm sw}}    
   P_{\hbox{\sevenrm sw}} + \gamma_{\hbox{\sevenrm pu}} 
   P_{\hbox{\sevenrm pu}}}{\rho_{\hbox{\sevenrm sw}}
       +\rho_{\hbox{\sevenrm pu}}} \;\; ,\nonumber
\end{eqnarray}
where $\gamma_{\hbox{\sevenrm sw}}$ ($\gamma_{\hbox{\sevenrm pu}}$) is
the ratio of specific heats for the solar wind (pickup ions).  This
exhibits an intuitive property, namely that the pressure terms can be
added in the numerator and densities can be added in the denominator,
imitating the situation for the spring constant and the mass in a
harmonic oscillator.  In general, \teq{\gamma_{\hbox{\sevenrm
pu}}\neq\gamma_{\hbox{\sevenrm sw}}}, and we observe that if the pickup
ions maintain the two-dimensional ring distribution of their injection,
as $\gamma =(d_{\hbox{\sevenrm pu}} + 2)/d_{\hbox{\sevenrm pu}}$, then
one would obtain $\gamma_{\hbox{\sevenrm pu}} =2$ for the pickup ion
$\gamma$.  Otherwise, if the pickup ions are isotropized, as will be
assumed later in the paper, then \teq{\gamma_{\hbox{\sevenrm pu}}
=5/3=\gamma_{\hbox{\sevenrm sw}}}.  It follows, that if
$d_{\hbox{\sevenrm pu}}$ and \teq{{\langle v^2\rangle}_{\hbox{\sevenrm
pu}}} are the dimensionality and mean square speed, respectively, of
the pickup ions, then
\begin{equation}
   {\cal M}_{\hbox{\sevenrm S}}\, =\, V_{\hbox{\sevenrm sk}} 
   \biggl\{  \dover{5\, n_{\hbox{\sevenrm sw}} k T/m}{
           3\, (n_{\hbox{\sevenrm sw}} + n_{\hbox{\sevenrm pu}})} +
     \dover{d_{\hbox{\sevenrm pu}} +2}{d_{\hbox{\sevenrm pu}}^2}\,
     \dover{n_{\hbox{\sevenrm pu}} {\langle v^2\rangle}_{\hbox{\sevenrm 
     pu}}}{n_{\hbox{\sevenrm sw}} +n_{\hbox{\sevenrm pu}} }\,\biggr\}^{-1/2} .
\end{equation}
At the termination shock, the solar wind is very cold, so that the
pickup ion component dominates the pressure, primarily because the
pickup ion abundance is significant (for protons, beyond around 5 AU,
the pickup ion density drops off roughly as \teq{1/r} since the
accumulated injection of pick up ions scales more or less as \teq{r},
which is diluted by the spherical expansion factor \teq{1/r^2}; this
contrasts the solar wind, whose density scales purely as the
\teq{1/r^2} dilution factor).  In this case,
\begin{equation}
   {\cal M}_{\hbox{\sevenrm S}} \;\approx\; \sqrt{
   \dover{d_{\hbox{\sevenrm pu}}^2}{d_{\hbox{\sevenrm pu}} + 2}\,
   \dover{n_{\hbox{\sevenrm sw}} + n_{\hbox{\sevenrm pu}}}{
          n_{\hbox{\sevenrm pu}}}\,
   \dover{V_{\hbox{\sevenrm sk}}^2}{{\langle v^2\rangle}_{\hbox{\sevenrm pu}}}
   } \;\; .
\end{equation}
It follows that the dependence of the sonic Mach number on the
dimensionality of the pickup ions is conveniently very weak, and that
pickup ion abundances \teq{n_{\hbox{\sevenrm pu}}/(n_{\hbox{\sevenrm sw}}
+ n_{\hbox{\sevenrm pu}})} exceeding around 1\%, limit the Mach
number to around ten.  As one moves from 1AU towards the termination
shock, adiabatic cooling of the solar wind forces \teq{{\cal
M}_{\hbox{\sevenrm S}}} to increase slowly ($\propto r$) until the
pickup ions dominate the pressure and the Mach number saturates at the
above value.

In all that follows, we take $d_{\hbox{\sevenrm pu}}=3$ and calculate
${\langle v^2\rangle}_{\hbox{\sevenrm pu}}$ directly from the injected
pickup distributions assuming they are isotropic in the local frame.

\subsection{Adiabatic Evolution of the Pickup Ion Distribution}

As the solar wind expands in its progression to the outer heliosphere,
it cools as does the pickup ion distribution.  The pickup ions are,
however, continually injected at rates depending on their distance from
the sun, so the determination of their distribution at the termination
shock is non-trivial.  The calculation of the injection rates and
resulting distribution function depends on details such as the radial
variations of the ionizing solar UV flux and solar wind density, and
the gravitational focusing of interstellar neutrals in the inner
heliosphere.  We use a standard expression for the pickup ion
phase-space distribution,  $f_{\hbox{\sevenrm pu}}(r,v)$, in the solar
wind frame, in the nose region of the termination shock (e.g.
Gloeckler et al.  1993, 1994; le Roux, Potgieter, \& Ptuskin  1996),
for a {\it three-dimensional} isotropic population, based on the
derivation of Vasyliunas \& Siscoe  (1976):
\begin{equation}
   f_{\hbox{\sevenrm pu}}(r,v) = \dover{3}{8 \pi}
   \left ( \dover{u_\infty}{V_{\hbox{\sevenrm sw}}^4} \right )\!
   \left ( \dover{\Lambda}{r} \right )\!
   \left ( \dover{v}{V_{\hbox{\sevenrm sw}}} \right )^{-3/2} \!\! n(r,\, v)
   \,\Theta(V_{\hbox{\sevenrm sw}} - v) \, ,
 \label{eq:PickupDist}
\end{equation}
where $v$ is the particle speed, $r$ is the radial distance (of the
termination shock in this application) from the sun along the line
pointing toward the nose of the termination shock, and $u_\infty \simeq
20$ km s$^{-1}$\ is the velocity of the Sun relative to the local
interstellar medium.  Here $\Lambda = \nu_{\rm E} r_{\rm E}^2/
u_\infty$ is the characteristic ionization distance for interstellar
neutrals, where $\nu_{\rm E}$ is the frequency of ionization at the
Earth, i.e.  at a  radial distance of \teq{r_{\hbox{\sevenrm E}}=1} AU
from the Sun.  While \teq{\Lambda} is written in terms of quantities
measured at 1 AU, it is actually independent of radius due to the
\teq{1/r^2} decline in the solar wind density (which is involved in
charge exchange with the interstellar neutrals) and the ionizing solar
UV flux.  The values we adopt for the ionization frequencies of various
ionic species at Earth are taken from the determinations at solar
minimum of Rucinski, Fahr, \& Grzedzielski (1993), and were those used
by le Roux, Potgieter, \& Ptuskin (1996), namely $\nu_{\rm E} = 5\times
10^{-7}$ s$^{-1}$, $6.7\times 10^{-8}$ s$^{-1}$, and $5\times 10^{-7}$
s$^{-1}$ for hydrogen, helium, and oxygen, respectively.  These values
differ significantly (at least for hydrogen and helium) from the
earlier values quoted by Vasyliunas \& Siscoe (1976) and fall below the
mean ionization frequencies recorded over the entire solar cycle by
factors of around 1.5 (e.g.  Rucinski et al.  1996).  The Heaviside
step function $\Theta (V_{\hbox{\sevenrm sw}} - v)$ is unity for
non-negative arguments and zero otherwise, so that it cuts the
distribution off at the solar wind speed, $V_{\hbox{\sevenrm sw}}$.
The factor $n(r,\, v)$ in Equation~(\ref{eq:PickupDist}) is given in terms of
the neutral density in the interstellar medium, $n_\infty$, by:
\begin{equation}
   n(r,\, v) = \dover{n_\infty}{4 \chi} (1 + \chi)^2
   \exp \left\{ -\left( \dover{\Lambda}{r} \right) \!
   \left( \dover{v}{V_{\hbox{\sevenrm sw}}} \right)^{-3/2} \!\!
   \dover{2}{\vert 1+ \chi \vert} \right\}  ,
 \label{eq:NeutralDen}
\end{equation}
with
\begin{equation}
   \chi^2 \; =\; 1 - \dover{r_p(0)}{r}
   \left( \dover{v}{V_{\hbox{\sevenrm sw}}} \right)^{-3/2} \;\; ,
\end{equation}
and
\begin{equation}
   r_p(0) \; =\; 2 G M_\odot (\mu - 1)/u_\infty^2 \;\; .
\end{equation}
In these expressions, $G$ is the gravitational constant, $M_\odot$ is
the mass of the Sun, and $\mu$ is the ratio of the solar radiation
pressure to the solar gravitational force.  Hence \teq{r_p(0)} is the
negative (for \teq{\mu\leq 1}) of the radius where the gravitational
potential (suitably modified for radiation pressure) equals the kinetic
energy of the interstellar neutrals, i.e.   approximately where
gravitational deflection of neutrals becomes important.  Note that,
except for the step function, the particle speed \teq{v} and the radial
distance \teq{r} in Eqs.~(\ref{eq:PickupDist})--(\ref{eq:NeutralDen})
always appear in the combination \teq{r_i\equiv (v/V_{\hbox{\sevenrm
sw}})^{3/2}\, r}, which is just the radius of injection of pickup ions
that adiabatically cool to speed \teq{v} at radius \teq{r}.  Therefore,
if \teq{n(r,\, v)} is expressed as a function of \teq{r_i}, $n$
represents the density of neutrals at the radius \teq{r_i} of
injection.  These characteristics of adiabatic evolution of the pickup
ion distribution were established by Vasyliunas \& Siscoe (1976), who
also presented results for two-dimensional pick-up ion populations.

Except for minor changes in notation, the above expressions are taken
directly from le Roux, Potgieter, \& Ptuskin  (1996), and following
them, we use $\mu = 0.7$ for hydrogen and $\mu = 0$ for helium and
oxygen to generate the pickup ion distributions.  For our comparisons
with the ACR observations presented in Cummings \& Stone (1996), we
normalize the pickup distributions generated with the above equations
to the densities estimated by Cummings and Stone.

\centerline{}
\vskip 0.2truecm
\centerline{\psfig{figure=apj99ejb_ts_f1.ps,width=8.8cm}}
\vskip 0.0truecm
\figcaption{
Upstream Phase-space densities for pickup ions expected at 85 AU
calculated using equations~(\ref{eq:PickupDist})
and~(\ref{eq:NeutralDen}) with $n_\infty({\rm H}) = 0.077$,
$n_\infty({\rm He}) = 0.01$, $n_\infty({\rm O}) = 9.7\times 10^{-5}$
cm$^{-3}$.  The velocity is in units of the solar wind speed,
$V_{\hbox{\sevenrm sw}}$. The three thermal ion species are all
injected with a temperature per nucleon of 535 K with charge states:
H$^+$, He$^{2+}$, and O$^{8+}$.  The pickup ions have a charge state of
$+1$.  The flat nature of the He$^{+}$\ distribution relative to those
of H and O reflects its much longer ionization length. 
   \label{fig:PickUpPhaseSpaceAll} }       
\centerline{}

\subsection{Direct Acceleration of Anomalous Cosmic Rays
at the Termination Shock}

We now present a model for the acceleration of anomalous cosmic ray
H$^{+}$, He$^{+}$, and O$^{+}$. Using the observations of Geiss et al.
(1994) and the model of Vasyliunas \& Siscoe  (1976) just discussed,
Cummings \& Stone  (1996) estimate the following pickup ion fluxes at
the nose of the heliosphere:  $F_{\hbox{\sevenrm p}}^{\hbox{\sevenrm
pu}} \simeq 1.0\times 10^{4}$ cm$^{-2}$~s$^{-1}$, $F_{\hbox{\sevenrm
He}}^{\hbox{\sevenrm pu}} \simeq 230$ cm$^{-2}$~s$^{-1}$, and
$F_{\hbox{\sevenrm O}}^{\hbox{\sevenrm pu}} \simeq 5.3$
cm$^{-2}$~s$^{-1}$.  Again assuming that the solar wind speed is
constant and equal to  500 km s$^{-1}$, the pickup densities at the
termination shock are then:  $n_{\hbox{\sevenrm p}}^{\hbox{\sevenrm
pu}} \simeq 2\times 10^{-4}$ cm$^{-3}$, $n_{\hbox{\sevenrm
He}}^{\hbox{\sevenrm pu}} \simeq 4.6\times 10^{-6}$ cm$^{-3}$, and
$n_{\hbox{\sevenrm O}}^{\hbox{\sevenrm pu}} \simeq 1.1\times 10^{-7}$
cm$^{-3}$.  These values correspond to $n_\infty(H) = 0.13$ cm$^{-3}$,
$n_\infty(He) = 0.02$ cm$^{-3}$, and $n_\infty(O) = 7\times 10^{-5}$
cm$^{-3}$, somewhat different from the values assumed by le Roux,
Potgieter, \& Ptuskin  (1996).  More recently, Gloeckler, Fisk, \&
Geiss  (1997) report  $n_\infty(H) = 0.115$ cm$^{-3}$\ and
$n_\infty(He) = 0.0153$ cm$^{-3}$.  These differences are relatively
small and we use the Cummings and Stone values to allow for a direct
comparison.  The values are listed in Table 1 under Model I along with
corresponding solar wind values at the Earth, for densities,
temperatures, and the estimated magnetic field strength.  We assume
$\Theta_{\rm Bn} = 89^\circ$, inject the thermal and pickup ions with
far upstream (i.e.  $-x\gg\eta r_{\rm g1}$, where $r_{\rm g1} \equiv
m_{\rm p} V_{\hbox{\sevenrm sk}} c/ e$) phase space
distributions as shown in Figure~\ref{fig:PickUpPhaseSpaceAll}, and use $\eta =
14$, chosen to give a good fit to the observed ACR intensities, as will
become evident shortly.


\centerline{}
\vskip 0.0truecm
\centerline{\psfig{figure=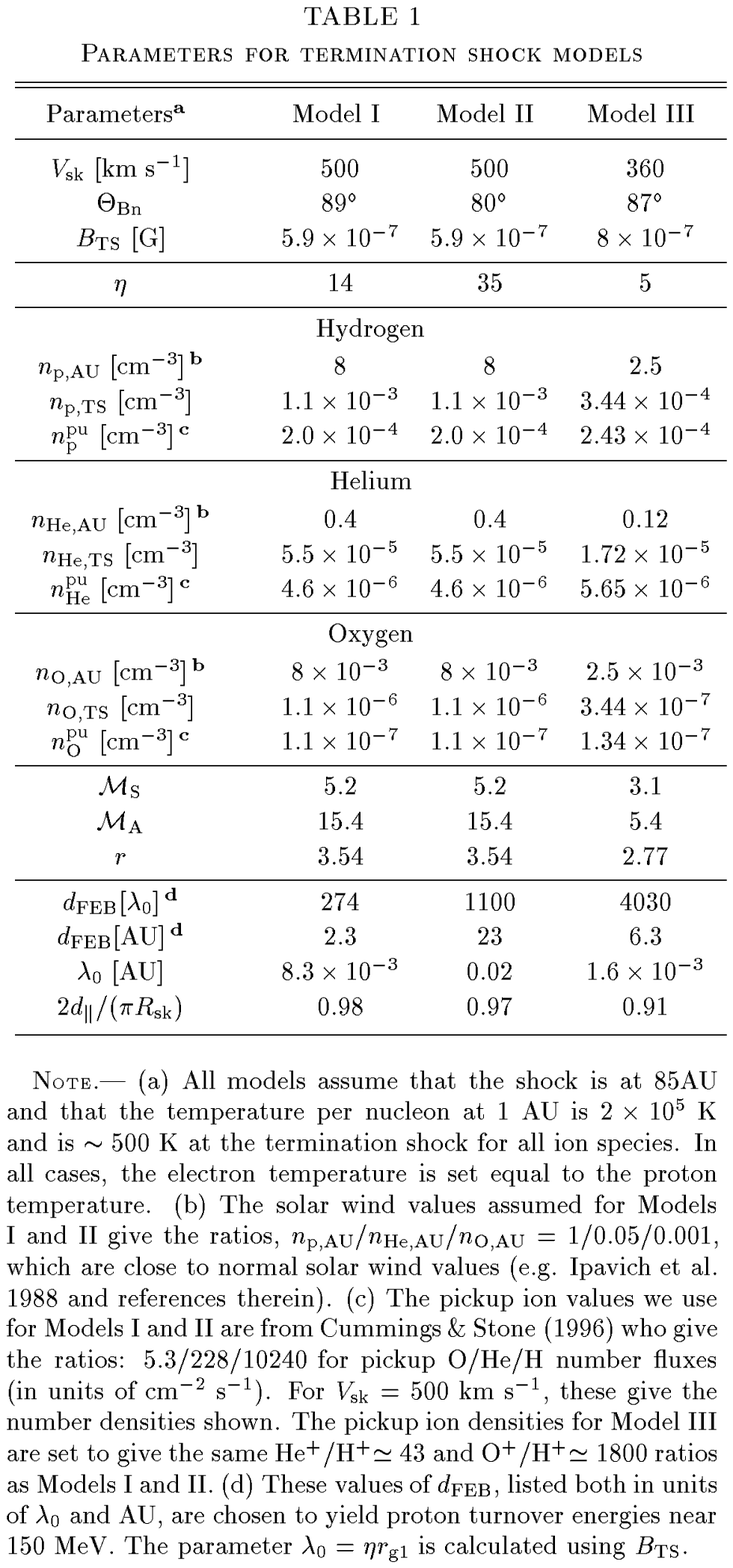,width=8.8cm}}
\vskip 0.0truecm
\centerline{}

\newpage

\centerline{}
\vskip 0.2truecm
\centerline{\psfig{figure=apj99ejb_ts_f2.ps,width=7.4cm}}
\vskip 0.0truecm
\figcaption{
Determination of shock structure by iteration.  The top panel shows the
$x$-component of the flow speed, $u_{\rm x}(x)$, the middle panel shows
the $xx$-component of momentum flux, and the bottom panel shows the
energy flux, all normalized to far upstream values.  In each panel, the
first and last iterations are shown as dashed lines and solid lines,
respectively.  The energy and momentum fluxes are conserved throughout
the shock to within 1\%.
   \label{fig:ProfPU} }       
\centerline{}

The self-consistent shock profile is shown as a solid  line in the top
panel of Figure~\ref{fig:ProfPU}, along with the $xx$-component of the momentum
flux and the energy flux in the lower two panels. In each panel, the
dashed line is the test-particle quantity obtained with the
discontinuous shock  and the solid line is the value obtained after the
self-consistent smooth shock structure has been found.  For a complete
description of how the shock structure is determined, see Ellison,
Baring, \& Jones (1996).  The important point is that,  even for an
obliquity of $89^\circ$, the injection and acceleration of thermal and
pickup ions is efficient enough to cause some departures from momentum
and energy conservation in the discontinuous shock; the downstream
fluxes rise to  a factor of $>1.1$ above the far upstream values.  Even
though the shock smoothing is quite small (Figure~\ref{fig:ProfPU} uses a
linear distance scale and the portion of the shock shown is a small
fraction of the size set by $d_{\hbox{\sevenrm FEB}}$), it is necessary
to conserve momentum and energy fluxes.  Our self-consistent, smooth
shock solution conserves all fluxes, including the $xz$-component of
momentum and uniformity of the tangential electric field (not shown),
across the shock.  The angle between the shock normal and the magnetic
field goes smoothly from $\Theta_{\rm Bn}=89^\circ$ far upstream to
$\Theta_{\rm Bn}=89.72^\circ$ downstream.  The addition of the pickup
ions has caused the sonic Mach number to decrease substantially from
the example we gave above, i.e.  here ${\cal M}_{\hbox{\sevenrm
S}}=5.2$ versus ${\cal M}_{\hbox{\sevenrm S}}=130$ without pickup ions
(see Table~1).

Note that even though $d_{\hbox{\sevenrm FEB}}$, which is measured
along the shock normal, may be a small fraction of the shock radius,
the high obliquity means that ions will move much greater distances
along the shock face.  Setting the distance parallel to the shock face
in our plane-shock approximation to be $d_{\parallel} \sim
d_{\hbox{\sevenrm FEB}} \tan{\Theta_{\rm Bn}}$, we require for
consistency that particles stream no more than the pole-to-equator
distance, i.e.
\begin{equation}
   d_{\parallel} \; <\; \pi R_{\rm sk} / 2\ ,
 \label{eq:dParlimit}
\end{equation}
or,
\begin{equation}
   d_{\hbox{\sevenrm FEB}} \;\lesssim\; 
   \dover{\pi R_{\rm sk}}{2 \tan{\Theta_{\rm Bn}}} \ .
\end{equation}
Clearly, the quasi-spherical geometry of the termination shock renders
the effective value of $d_{\parallel}$ somewhat (but not much) less
than the bound in equation~(\ref{eq:dParlimit}).  The quantity, $2
d_{\parallel} / (\pi R_{\rm sk})$ is listed in Table 1 and for this
example (Model I), $2 d_{\parallel} / (\pi R_{\rm sk}) \simeq 1.1$.

In Figure~\ref{fig:PickUpSpecAll} we show the model spectra, calculated at the
termination shock, along with Voyager 1 (V1) data measured (well within
the termination shock)  during 1994 on days 157-313  (Cummings \& Stone
1996; see also Christian, Cummings, \& Stone 1995).  The value of
$\eta$ has been chosen to obtain general agreement with the
normalization of the ACR proton observations, but there has been no
other  adjustment of normalization in the top panel.  Smaller values of
$\eta$ (i.e.  stronger scattering) would yield higher model intensities
at ACR energies (i.e.  in conflict with the data) and larger values
would yield lower model intensities. Note that the intensity of the ACR
H$^{+}$\ peak at $\sim 50$ MeV is $\sim 10$ orders of magnitude below
the H$^{+}$\ pickup bump at $\sim 2$ keV; only a tiny fraction of
pickup ions need to be accelerated to energies above $\sim 10$ MeV to
account for the observed ACR fluxes.  The value of \teq{\eta =14} used
to obtain the ``fit'' should be regarded only as a rough indication of
the true value, given the sensitivity of the ACR flux to the pick-up
ion abundance and the shape of their distribution.

In examining Figure~\ref{fig:PickUpSpecAll} it must be remembered that the V1
observations were made at an average radial location of 57 AU and show
the effects of solar modulation.  Our model spectra, on the other hand,
are calculated at the termination shock and do not include modulation.
Note that our shock acceleration simulation does generate low energy
``modulation-like'' depletions in upstream populations (e.g.  Baring,
Ellison, \& Jones 1994; Ellison, Baring, \& Jones 1996) due to
inefficient convection against the fluid flow, much like the model of
Lee (1982).  However, for the termination shock models of this paper,
such depletions appear relatively close to the shock, on scales $\ll
D_{\hbox{\sevenrm TS}}$, due to our incomplete modeling of particle
convection and diffusion in the complex geometry of the heliosphere.
The heavy solid line in Figure~\ref{fig:PickUpSpecAll} is the estimate of
Cummings \& Stone  (1996) for the power law H$^{+}$\ spectrum at the
termination shock and, as mentioned above, we have chosen $\eta$ to
approximately match this intensity (fine tuning of $\eta$ would give a
more precise match). 

\centerline{}
\vskip 0.2truecm
\centerline{\psfig{figure=apj99ejb_ts_f3.ps,width=8.8cm}}
\vskip 0.0truecm
\figcaption{
Comparison of Voyager 1 observations of ACR H, He, and O (made during
1994/157-313 when V1 was at an average radial location of $\sim 57$ AU)
to Model I spectra calculated {\it at the termination shock}. The model
spectra have an absolute normalization determined by the injection
parameters, i.e.  $n_{\hbox{\sevenrm p,TS}} V_{\hbox{\sevenrm sw}} =
5.5\times 10^{4}$ cm$^{-2}$ s$^{-1}$ \ for the protons and
corresponding values for the He and O. The value of $\eta$ has been
chosen to give a general fit to the intensities of the observed ACR's.
The sharp thermal peaks show the relatively cold solar wind ions that
have not yet thermalized. As the observation position is moved
downstream, these peaks broaden.  Note that the H thermal peak
intensity is $\sim 11$ orders of magnitude above the observed ACR
intensity. The heavy solid line is the Cummings \& Stone (1996)
estimate for the ACR proton intensity at the termination shock.  In the
bottom panel, we have individually adjusted the normalizations to match
the ACR observations. The relative adjustments for He$^{+}$\ and
O$^{+}$\ are labeled.
   \label{fig:PickUpSpecAll} }       
\centerline{}

There are several points to consider.  First, the limits on the maximum
ACR energy are such that the Cummings \& Stone extrapolation extends
into the exponential cutoff.  Second, even though the shock model we
are using has a compression ratio of $r=3.54$, well above that inferred
by Cummings \& Stone, the spectral {\it slope} in the very limited
energy range (i.e.  above the modulation turnover) provided by the data
is reasonably well fit.  Because of the limited energy range and the
spectral cutoff, it may not be possible to meaningfully constrain the
termination shock Mach number by extrapolating ACR observations made
well inside the heliosphere back to the termination shock as done by
Cummings \& Stone  (1996).  Third, the ACR data clearly show that the
observed He$^{+}$/H$^{+}$\ ratio is greater than our model predicts.
This is also true for the O$^{+}$/H$^{+}$\ ratio and in order to match
the observed fluxes, we would have to increase, relative to H, the
He$^{+}$\ intensity by a factor $\sim 4$, and the O$^{+}$\ intensity a
factor $\sim12$.  This adjustment has been done in the bottom panel of
Figure~\ref{fig:PickUpSpecAll} to show that the shapes of the observed ACR
spectra are matched exceeding well by our model above the modulation
turnover. In particular, our single parameter $d_{\hbox{\sevenrm
FEB}}$, simultaneously gives a good match to the cutoff for all three
species.

Considering the uncertainties involved in estimating the various
parameters needed at the termination shock, such as the solar wind
speed and the pickup ion densities, we believe the match indicated in
Figure~\ref{fig:PickUpSpecAll} is acceptable.

\centerline{}
\vskip 0.2truecm
\centerline{\psfig{figure=apj99ejb_ts_f4.ps,width=8.8cm}}
\vskip 0.0truecm
\figcaption{
Acceleration time in years versus energy per nucleon for the
H$^{+}$\ (solid line), He$^{+}$\ (dashed line), and
O$^{+}$\ (dot-dashed) produced in Model I. These curves are calculated
from the analytic result given in Ellison, Baring, \& Jones  (1995).
The upper limits on the O acceleration time from electron stripping at
10 and 20 MeV/A are from Adams \& Leising (1991).  The \teq{70^\circ},
\teq{\eta =25} curve is included to indicate what shock parameters are
necessary to encroach upon the experimental upper limits.
   \label{fig:AccTime} }       
\centerline{}

\subsubsection{Acceleration Time}

An important constraint on the production of ACRs comes from the charge
stripping rate; clearly considerations of ACR generation are simplified
when charge-stripping timescales exceed those of acceleration.  Such
ionization is relevant to species heavier than He (whose stripping
timescales are long), in this case oxygen.  Adams \& Leising (1991)
showed that 10 MeV/A singly charged oxygen ions will be further
stripped, in conflict with observations, if they propagate more than
$\sim 0.2$ pc in the local interstellar medium. Jokipii (1992) showed
how this relates to the acceleration rates of various mechanisms, and
concluded that first-order Fermi acceleration at highly oblique shocks
is the only mechanism fast enough to satisfy this limit.  Our results
are in agreement with this assessment and we plot the acceleration time
versus energy per nucleon in Figure~\ref{fig:AccTime} for our Model I. Here,
the dot-dashed line is O$^{+}$, the dashed line is He$^{+}$, and the
solid line is H$^{+}$.  The plots in Figure~\ref{fig:AccTime} were calculated
using the analytic result of Ellison, Baring, \& Jones  (1995) [i.e.
equation~(4) of that paper with $E_i = 0.6$ MeV/A], but our direct
Monte Carlo  determination of the acceleration time  is consistent with
this at superthermal energies, as also shown in Ellison, Baring, \&
Jones  (1995).

The actual limits of 4.6 and 6.3 yrs placed on the propagation of
oxygen by Adams \& Leising (1991) are shown as upper limits at 10 and
20 MeV/A, respectively.  The acceleration time is well below these
limits, though we also depict a \teq{70^\circ}, \teq{\eta =25} case to
indicate what type of shock parameters might be needed for charge
stripping to be relevant.  In addition to the limits of Adams \&
Leising, there is also the report of observations of ACR oxygen in
higher ionization states than O$^{+}$\ (Mewaldt et al. 1996), a
constraint that provides a lower limit to the diffusive acceleration
timescale.  Given that such energies per nucleon are at the upper end
of the oxygen spectrum in the models presented here (e.g.  see
Figure~\ref{fig:SpecFlat} below), it appears that detailed consideration of ACR
propagation and diffusion in the heliosphere is necessary to obtain a
suitable description of energetic oxygen in various ionization states.
As mentioned above, we have not included charge stripping in our
present calculation, but will include this effect in future work.  We
remark that Jokipii (1996) has included electron stripping in his
acceleration and transport model and finds good agreement with these
observations.

\subsection{Limited Parameter Survey}

\vskip-12pt
\subsubsection{Variation of $\Theta_{\rm Bn}$}

It is instructive to explore how our model output changes with
variations in its most important parameters, namely $\eta$ and
$\Theta_{\rm Bn}$. In Model II we have changed $\Theta_{\rm Bn}$ from
$89^\circ$ to $80^\circ$ to see the effect this has on our fits to the
ACR observations. We have kept all other input parameters the same as
in Model I except that we have altered $\eta$ to give a general fit to
the H ACR intensity as before.  The injection efficiency depends
strongly on both $\eta$ and  $\Theta_{\rm Bn}$ (e.g.  Ellison, Baring,
\& Jones  1995), increasing as either $\Theta_{\rm Bn}$ or $\eta$ is
decreased. By decreasing $\Theta_{\rm Bn}$ from $89^\circ$ to
$80^\circ$, we must reduce the scattering efficiency (in this case by
setting $\eta=35$) to obtain a fit to the ACR H$^{+}$\ intensities.
Once this adjustment is made, the  characteristics of the $80^\circ$
and $89^\circ$ results are similar as is shown in Figure~\ref{fig:ThreeSpec},
where we compare the proton spectra from Model I (solid line)  and the
$80^\circ$ example, Model II (dashed line).  While $\eta$ has changed
compared to Model I, the maximum energy has been kept essentially the
same by varying $d_{\hbox{\sevenrm FEB}}$.  Of course we do not answer
(or even address) the question of how the magnetic turbulence is
produced, or why it obtains a level which gives observed ACR
intensities (i.e.  why $\eta$ has a particular value).  However, this
example does show that the injection process is perfectly well defined
within standard diffusive shock acceleration and that a smooth change
in parameters results in a continuous change in output efficiencies.

\centerline{}
\vskip 0.2truecm
\centerline{\psfig{figure=apj99ejb_ts_f5.ps,width=8.8cm}}
\vskip 0.0truecm
\figcaption{
Comparison of the proton spectra for Models I (solid line), II (dashed
line), and III (dotted line), illustrating the variation of model
output with shock obliquity (comparing Models I and II) and shock
strength (Model I vs. III), keeping the 10 MeV/A ACR flux more or less
constant by adjusting the value of \teq{\eta}.  The heavy solid line is
the Cummings \& Stone  (1996) estimate for the ACR proton power-law
intensity at the termination shock. Fine tuning of $\eta$ would allow a
more exact match between our models and the Cummings \& Stone
intensity.
   \label{fig:ThreeSpec} }       
\centerline{}

It is also true that if  $\Theta_{\rm Bn}$ is decreased much below
$80^\circ$, the value of $\eta$ required to obtain the observed ACR
intensity will become so large that the size of the foreshock region is
greater than the termination shock radius or that  $d_{\parallel} > \pi
R_{\rm sk} / 2$.  Note from Table 1 that $d_{\hbox{\sevenrm FEB}}$ goes
from 2.3  AU for Model I to 23 AU for Model II.  This suggests that, at
least within the simple assumptions we have made here, the termination
shock cannot  be injecting and accelerating pickup ions for significant
times in states where the local $\Theta_{\rm Bn} \lesssim 70^\circ$. In
order to match the ACR intensities, the increased injection resulting
from the low obliquity must be matched by a decrease in scattering
efficiency which implies length scales which are inconsistent with the
size limitations of the termination shock. It may be possible, however,
that large departures from highly oblique conditions last for short
times (e.g.  Kucharek \& Scholer 1995), but if these conditions result
in enhanced injection, as has been suggested, the time-averaged $\eta$
must be correspondingly increased to satisfy the observed ACR
intensities.  The time needed to accelerate ions to ACR energies will
also increase as $\Theta_{\rm Bn}$ is decreased and $\eta$ is
increased. The dotted line in  Figure~\ref{fig:AccTime} shows the acceleration
time for O$^+$ when $\Theta_{\rm Bn}=70^\circ$ and $\eta=25$.  While
this is still consistent with the upper limits of Adams \& Leising
(1991), it does suggest that a much larger fraction of ACRs will be
multiply charged if non-highly oblique portions of the termination
shock contribute significantly to the observed ACRs.

\subsubsection{Low Mach Number Example and Effect of Pickup Ions}

The models we have used so far have all had sonic Mach numbers  ${\cal
M}_{\hbox{\sevenrm S}} \gtrsim 5$ and compression ratios $r \gtrsim
3.5$. These compression ratios are considerably larger than the $r\sim
2.6$ estimated by Stone, Cummings, \& Webber (1996) and Cummings \&
Stone  (1996) from the ACR spectral shapes, but they are what would be
expected for a solar wind speed of $\sim 500$ km s$^{-1}$\ and the
densities estimated by Cummings \& Stone.  To investigate the effect of
Mach number, we have performed another simulation where we have
modified  our parameters to yield a weaker shock, i.e.  $r\simeq 2.8$.
We use a smaller solar wind speed at the termination shock, i.e.
$V_{\hbox{\sevenrm sk}} =360$ km s$^{-1}$\  (as estimated by  Isenberg
1997), and have adjusted our solar wind and pickup ion densities and
other parameters (e.g.  $\Theta_{\rm Bn} = 87^\circ$) as indicated by
Model III in Table 1, to maintain $d_{\parallel}\sim \pi R_{\rm
sk}/2$.  As before, we iterate to a self-consistent shock structure
after adjusting $\eta$ to give a reasonable fit to the ACR intensity.

\centerline{}
\vskip 0.2truecm
\centerline{\psfig{figure=apj99ejb_ts_f6.ps,width=8.8cm}}
\vskip 0.0truecm
\figcaption{
Same as Figure~\ref{fig:PickUpSpecAll} for Model III.  
As in Figure~\ref{fig:PickUpSpecAll}, in the bottom panel we have individually
adjusted the normalizations to match the ACR observations.  The
relative adjustments for He$^{+}$\ and O$^{+}$\ are labeled.
   \label{fig:ACR_g_Five} }       
\centerline{}

The low compression ratio produces a steeper spectrum than in our
previous examples, and in order to match the ACR intensity at $\sim
100$ MeV, a larger injection efficiency (i.e.  smaller $\eta$) is
required. We find that $\eta \simeq 5$ yields a good match to the ACR
observations as shown in Figure~\ref{fig:ACR_g_Five}. Any small discrepancies
between this model (or the others) and the ACR H$^{+}$\ intensity
(extrapolated by Cummings \& Stone) can be removed by fine tuning
$\eta$.  The difficulty in deducing the shock strength from the
spectral shape (which is strongly influenced by the non-linear shock
smoothing) in the limited energy range afforded by the ACRs is also
obvious from this Figure.

\centerline{}
\vskip 0.2truecm
\centerline{\psfig{figure=apj99ejb_ts_f7.ps,width=8.8cm}}
\vskip 0.0truecm
\figcaption{
Spectra from Models I and III renormalized and multiplied by
$(E/A)^{1.5}$. In each case, we have normalized all spectra to the same
pickup ion density, i.e.  for Model I we have multiplied the He
spectrum by $n_{\hbox{\sevenrm p}}^{\hbox{\sevenrm
pu}}/n_{\hbox{\sevenrm He}}^{\hbox{\sevenrm pu}} \simeq 43$ and the
oxygen by $n_{\hbox{\sevenrm p}}^{\hbox{\sevenrm pu}}/n_{\hbox{\sevenrm
O}}^{\hbox{\sevenrm pu}} \simeq 1800$, and for Model III, we have
multiplied the He spectrum by $n_{\hbox{\sevenrm p}}^{\hbox{\sevenrm
pu}}/n_{\hbox{\sevenrm He}}^{\hbox{\sevenrm pu}} \simeq 170$ and the
oxygen by $n_{\hbox{\sevenrm p}}^{\hbox{\sevenrm pu}}/n_{\hbox{\sevenrm
O}}^{\hbox{\sevenrm pu}} \simeq 7000$.  In the top two panels, the
self-consistent smooth shock is used to produce the spectra and a clear
$A/Q$ enhancement of He$^{+}$\ or O$^{+}$\ to H$^{+}$\ is seen.  In the
bottom panel, we determined the spectra using the test-particle,
discontinuous shock and essentially no enhancement (other than
statistical variations) is present.
   \label{fig:SpecFlat} }       
\centerline{}

It has been known for some time that the acceleration efficiency of
shocks that are smoothed by the pressure of accelerated particles is an
increasing function of $A/Q$ (e.g.  Eichler 1979; Ellison, Jones, \&
Eichler  1981; see Ellison, Drury, \& Meyer  1997 for a recent
reference) in quasi-parallel scenarios.  This effect, which depends
only on the conservation of momentum and a spatial diffusion
coefficient which is an increasing function of energy, occurs because
non-relativistic ions with larger $A/Q$ (i.e.  larger rigidities) have
longer upstream diffusion lengths, at a given energy per nucleon. The
fact that the shock is smoothed means that the high $A/Q$ particles
`feel' a larger effective compression ratio and are accelerated more
efficiently and, the greater the smoothing, the greater the
enhancement.  Enhancements have been confirmed at the quasi-parallel
Earth bow shock (i.e.  Ellison, M\"obius, \& Paschmann  1990) and
should occur regardless of the shock obliquity {\it as long as the
shock is smoothed}.  In order to investigate this $A/Q$ enhancement, we
re-plot the model spectra from Figures~\ref{fig:PickUpSpecAll} and
\ref{fig:ACR_g_Five}, renormalizing the helium and oxygen spectra so
they have the same upstream pickup ion number density as hydrogen, so
that any difference produced during acceleration can be seen directly.
That is, for both Models  I and III, we multiply the helium by
$n_{\hbox{\sevenrm p}}^{\hbox{\sevenrm pu}}/n_{\hbox{\sevenrm
He}}^{\hbox{\sevenrm pu}} \simeq 43$ and the oxygen by
$n_{\hbox{\sevenrm p}}^{\hbox{\sevenrm pu}}/n_{\hbox{\sevenrm
O}}^{\hbox{\sevenrm pu}} \simeq 1800$.

\centerline{}
\vskip 0.2truecm
\centerline{\psfig{figure=apj99ejb_ts_f8.ps,width=8.8cm}}
\vskip 0.0truecm
\figcaption{
Acceleration efficiency in terms of 
the fraction of energy density in ions with energy per nucleon $E/A$
and above. The solid, dashed, and dot-dashed curves are the H, He, and
O efficiencies determined from Model I. Intercepts with the horizontal
line show the energy per nucleon where each species is 1\% efficient.
   \label{fig:EnEffPU} }       
\centerline{}

The results are shown in the top two panels of Figure~\ref{fig:SpecFlat} with
all spectra multiplied by $(E/A)^{1.5}$.  Both Models show an $A/Q$
enhancement effect and although it is somewhat larger in Model III than
Model I, it is still not as strong as deduced by Cummings \& Stone
(1996).  The bottom panel of Figure~\ref{fig:SpecFlat} shows Model III with no
shock smoothing but all other parameters the same. Here, there is
essentially no difference in the various spectra, other than at the
high energy turnover, as expected.  As indicated in the lower panel of
Figure~\ref{fig:ACR_g_Five}, our Model III He$^{+}$/H$^{+}$\ and
O$^{+}$/H$^{+}$\ ratios are still lower than the observed ACR ratios by
a factor of $\sim 5$.  The actual acceleration efficiency
$\epsilon(E/A\! >)$ for Model I, as defined above to be the fraction of
energy density in particles of energy per nucleon $E/A$ and above, is
shown in Figure~\ref{fig:EnEffPU}. From this we see that 1\% of the energy
density (horizontal line) lies above $\sim 10$ keV/A for all three
species.  Note that at high energy per nucleon, the protons dominate
(see also Figure~\ref{fig:SpecFlat}) because they extend to higher $E/A$ for a
given gyroradius.

\centerline{}
\vskip 0.2truecm
\centerline{\psfig{figure=apj99ejb_ts_f9.ps,width=8.8cm}}
\vskip 0.0truecm
\figcaption{
Shock profiles [i.e. $u(x)/V_{\hbox{\sevenrm sk}}$] for models I, II,
and III.  Note that the different values of $u(x)/V_{\hbox{\sevenrm
sk}}$ in the downstream region (i.e.  $x > 0$) result from different
compression ratios.  The abscissa is scaled in units of the upstream
diffusion length (times 3) for protons travelling with the solar wind
speed (see Equation~[\ref{eq:diffscale}]).  The arrows at the bottom
denote the different {\it upstream} scalelengths for diffusion of the
three pickup species, obtained by setting \teq{v=V_{\hbox{\sevenrm
sw}}} in Equation~(\ref{eq:diffscale}).
   \label{fig:ProfThree} }       
\centerline{}

The somewhat larger $A/Q$ enhancement of Model III compared to Model I
may arise due to the different shock structures for these models; these
are exhibited in Figure~\ref{fig:ProfThree}.  In comparing smoothing in the
various models, it is instructive to scale the distance normal to the
shock in units of the diffusion length in the normal direction.  For
field obliquity \teq{\Theta_{\rm Bn}}, the diffusion coefficient in
this direction is \teq{\kappa_{xx}=\kappa_{\parallel}\cos^2\Theta_{\rm
Bn} +\kappa_{\perp}\sin^2\Theta_{\rm Bn}=\kappa_{\parallel}
[\cos^2\Theta_{\rm Bn} +\sin^2\Theta_{\rm Bn}/(1+\eta^2)]} using
kinetic theory to relate \teq{\kappa_{\perp}} to
\teq{\kappa_{\parallel}}.  It then follows, using the scaling units of
\teq{r_{\rm g1}=m_pV_{\hbox{\sevenrm sw}} c/e}, that the diffusion
length \teq{\kappa_{xx}/V_{\hbox{\sevenrm sw}}} is given by
\begin{equation}
   \dover{\kappa_{xx}}{V_{\hbox{\sevenrm sw}}} \,=\, \dover{1}{3}\,
   \dover{A}{Q} \left( \dover{v}{V_{\hbox{\sevenrm sw}}} \right)^2
   \eta r_{\rm g1}\;  \left\{ \cos^2\Theta_{\rm Bn}
                      +\dover{\sin^2\Theta_{\rm Bn}}{1+\eta^2}\right\}
 \label{eq:diffscale}
\end{equation}
for particles of speed \teq{v}.  The factor in curly brackets times
\teq{\eta r_{\rm g1}} is used as the length unit in Figure~\ref{fig:ProfThree}.
Hence the Figure gives an indication of the relative smoothing incurred
in the different models.  Model I is smoother than Model III which
seems in conflict with the fact that Model III shows a larger $A/Q$
effect.  This behavior indicates the complexity of such highly oblique
systems, which will depend on other factors such as the shock speed,
Mach number, the total compression ratio, and pick-up ion abundances.
Furthermore, the spectra in Figure~\ref{fig:SpecFlat} indicate that Model III
is a more efficient injector than Model I, a property which follows
from the sharper nature of the Model III profile.  In
Figure~\ref{fig:ProfThree}, arrows mark the typical upstream diffusion scales of
the three pick-up ion species (relative to the shock), determined by
setting \teq{v=2V_{\hbox{\sevenrm sw}}} in Equation~(\ref{eq:diffscale}).
These will be somewhat modified in the downstream region due to the
different field obliquity there.  The diffusion lengths indicate that
little \teq{A/Q} enhancement would be expected for Model II, and that
most should be seen for Models I \& III, given that helium and oxygen
pickup ions sample much larger compression ratios than hydrogen in
these two cases.  The interpretation of the $A/Q$ enhancement is
further complicated by the fact that diffusion in the downstream region
(whose scales are not exhibited in the Figure) modifies the typical
scalelengths, and that this depends in a complicated manner on the
values of \teq{\Theta_{\rm Bn}}, \teq{\eta}, the Mach number, and the
overall compression ratio.  We remark that such complexities of
\teq{A/Q} enhancement behavior are diminished in strong shocks and
particularly in quasi-parallel ones, where the number of influential
shock parameters is reduced.

\centerline{}
\vskip 0.2truecm
\centerline{\psfig{figure=apj99ejb_ts_f10.ps,width=8.8cm}}
\vskip 0.0truecm
\figcaption{
The solid line is the same proton spectrum shown in Figure~\ref{fig:ACR_g_Five},
as are the ACR proton data and the Cummings and Stone extrapolation.
The dotted and dot-dashed lines are calculated with no pickup ions for
the two values of $\eta$ shown. Injection is extremely sensitive to
$\eta$, but for scattering at the Bohm limit ($\eta = 1$), ACR
intensities can be produced at the termination shock with only thermal
solar wind ions.
   \label{fig:NoPickUp} }       
\centerline{}

Even though we have had to make some  changes in  Model III from our
previous examples to reduce $2 d_{\parallel} / (\pi R_{\rm sk})$ to
$\sim 1$  (we have increased $B_{\hbox{\sevenrm TS}}$ to $8\times
10^{-7}$ G and lowered $\Theta_{\rm Bn}$ to $87^\circ$), this model, as
well as our others, has reasonable values for the important
parameters.  Hence, we believe that our results describe the global
qualitative properties of ACR acceleration at the termination shock, so
that only minor fine-tuning is necessary if a more accurate data/theory
comparison is desired.

Finally, we note that pickup ions are not absolutely necessary for
producing the ACRs.  The dot-dashed and dotted lines in
Figure~\ref{fig:NoPickUp} show proton spectra produced for Model III when {\it
only thermal protons} are injected at the termination shock. The effect
the superthermal pickup ions have on the overall acceleration
efficiency is dramatic in the case where $\eta=5$ (without them the
intensity at ACR energies drops by $\sim 10$ orders of magnitude), but
for stronger scattering, the ACR intensities could be produced solely
from thermal ions. The dotted curve shows the spectrum produced,
without pickup ions, assuming $\eta=1$, i.e.  the Bohm limit.  While
pickup ions are clearly dominant in the production of ACRs (charge
states unambiguously show this), it is important to note that some
acceleration of thermal ions does occur and the relative importance
will depend on the strength of scattering.

\section{Discussion}

\subsection{Issues Concerning Ion Injection}

Various proposals for a pre-injection acceleration phase at highly
oblique shocks have been put forward in the literature, a number of
which were mentioned in the Introduction.  While these hypotheses may
have stemmed from many reports of ``injection problems'' for ACRs, and
may indeed arise at the termination shock, the results of this paper
have shown that the solar wind termination shock can easily inject and
accelerate pickup ions to anomalous cosmic ray energies and intensities
if standard diffusive shock acceleration operates.  Hence, we find that
{\it no pre-injection stage is necessary}; the only requirement for
injection and acceleration of pickup ions consistent with ACR
observations is that strong enough magnetic turbulence be present near
the termination shock (i.e.  $\lambda_{\parallel} \sim 10 \, r_{\rm
g}$, implying $\kappa_{\perp} /  \kappa_{\parallel} \sim 0.01$) to
produce cross-field diffusion.  Self-generated turbulence of this
strength or greater is seen or inferred near a host of other
astrophysical shocks, including highly oblique interplanetary shocks
with parameters not too different from what is expected at the
termination shock (e.g.  see the recent analysis of {\it in situ}
Ulysses observations by Baring et al.  1997).

In this paper we have also shown that while the injection efficiency
depends fairly strongly on the shock obliquity and $\eta$, the
character of the injection does not and varies smoothly over a range of
parameters.  Furthermore, small changes in the shape of the pickup ion
distribution produce no noticeable effect on the injection and
acceleration efficiencies.  Our results seem consistent with the fact
that all directly observed collisionless shocks, with the sole
exception of the highly oblique Earth bow shock, accelerate thermal
ions directly and diffusively with reasonable efficiencies.  At the
quasi-perpendicular bow shock, the unique geometry (where the solar
wind constantly sweeps the magnetic field and particles past the
relatively tiny tangent point) prevents self-generated turbulence from
forming in the quasi-perpendicular precursor and readily explains why
particle acceleration there is restricted to reflected beams (e.g.
Ipavich 1988).  In highly oblique interplanetary shocks on the other
hand, the geometry is quite different (and similar to that expected at
the termination shock), with injection being effected on quite small
spatial scales and diffusive particle injection and acceleration
readily occurs even for thermal solar wind particles (e.g.  Baring et
al. 1997).

Particle injection at oblique shocks is, in fact, more difficult than
at parallel ones (e.g. Jokipii 1987) because downstream shock heated
particles have a harder time returning to the shock;  they move largely
along the oblique field lines if scattering is weak. For injection to
be efficient without energetic seed particles, particularly at  high
Mach numbers, the scattering must be reasonably strong and cross-field
diffusion must take place for injection to occur (see Ellison, Baring,
\& Jones  1996 for a general discussion of injection efficiency in
oblique shocks). Background magnetic turbulence in the undisturbed
solar wind appears not to be strong enough to provide this cross-field
diffusion, and it is not obvious that the termination shock can
generate enough local magnetic turbulence to produce it. This has led
to computer plasma simulations analogous to the Quest (1988) work on
parallel shocks, and these studies thus far have suggested that pickup
ions {\it cannot} be injected directly at the quasi-perpendicular
termination shock. For instance, Kucharek \& Scholer (1995) obtained
results with a one-dimensional hybrid simulation that showed {\it no
injection} of pickup ions for $\Theta_{\rm Bn} \gtrsim 60^\circ$.
Similar results were obtained by Liewer, Rath, \& Goldstein 1995.
Unfortunately, because of the extreme computing requirements  of
three-dimensional simulations, the self-consistent hybrid simulations
have so far been done mainly in restricted dimensionality.  Jokipii,
K\'ota, \& Giacalone  (1993) showed (see also the more detailed
derivation of Jones, Jokipii, \& Baring 1998, and also Giacalone \&
Jokipii 1994 and Giacalone 1994 for simulation work) that the presence
of an ignorable coordinate results in an artificial suppression of
cross-field transport. Thus, the essential physics needed for injection
has not been modeled correctly in the self-consistent one- and
two-dimension hybrid simulations so far applied to the termination
shock. This may also have led to the assertion that a pre-injection
stage is necessary when, in fact, full three-dimensional hybrid
simulations (run long enough, with enough particles, in a large enough
simulation box to allow for the development of mature turbulence and
particle acceleration) are required to definitively answer this
question.

We note that when ad hoc scattering, which allows cross-field
scattering, is added to a one-dimensional  hybrid simulation
(Giacalone, Jokipii, \& K\'ota 1994), particle injection does take
place at perpendicular shocks.  These simulations are still severely
restricted in dynamic range and cannot produce energies typical of
ACRs, but as far as we can tell, they do see the beginnings of
injection and seem consistent with our results as far as a comparison
can be made.  {\it It must be emphasized that the only injection
problem that exists for quasi-perpendicular shocks  is whether or not
magnetic turbulence of the required wavelengths to interact with shock
heated  ions is strong enough to produce cross-field diffusion. } If it
is, we know of nothing in the Fermi mechanism that will prevent
injection and acceleration.

\vskip 13pt

\subsection{Comparison with Other Models of ACR Production}

A number of models of ACR acceleration have been presented which solve
numerically the so-called Parker transport equation (e.g.  Parker 1965)
or similar kinetic equations which require near-isotropic
distributions.  Jokipii and co-workers (e.g.  Jokipii \& Giacalone
1996) solve the full equation in two-dimensions for a spherical
termination shock and follow the acceleration of superthermal particles
(i.e.  $\gtrsim 100$ keV/nuc) in a realistic solar wind configuration.
They include the Parker spiral magnetic field, curvature and gradient
drifts, adiabatic losses, charge stripping, an equatorial current
sheet, and 11-year sunspot cycle magnetic field reversals.  The
superthermal particles are injected as test-particles and their
distribution function is followed during acceleration and propagation
to an observation point in the inner heliosphere.  The turnover of the
ACR spectra near 150 MeV/A comes naturally in this model from the
potential drop between the pole and the equator and only depends on the
rotation rate of the sun and the magnetic-field strength.  In all of
the above respects, except for treating the accelerated particles as
test-particles and starting ACRs off as mildly energetic rather than at
solar wind or pickup ion energies to ensure their distributions are
nearly isotropic, the Jokipii model is more complete than ours and has
been successful in modeling ACR spectral shapes (including multiply
charged ACRs; Jokipii 1996), latitudinal gradients, and other aspects
of solar modulation.

Le Roux, Potgieter, \& Ptuskin (1996) (see also Le Roux \& Fichtner
1997) investigate the acceleration and modulation of ACRs including the
modification of the termination shock from the pressure of the ACRs as
well as galactic cosmic rays. They solve the transport equation and
determine the shock structure with a set of time-dependent conservation
equations. While this model is quite advanced, they obtain multiple
solutions (i.e.  le Roux \& Fichtner 1997) with quite different values
for their free injection parameter.  Chalov \& Fahr (1996a) present a
so-called three-fluid model (solar wind plasma, pickup ions, and ACRs)
which also yields the shock structure under the influence of ACR
acceleration.  Again, as with all fluid models of shock structure,
injection is treated parametrically and all results depend critically
on the injection parameter.

In contrast, our approach has concentrated on the injection process and
the self-consistent determination of the shock structure in the
plane-shock approximation, assuming that other aspects, such as a
realistic geometry and detailed propagation models, have a lesser
effect on the observed ACR spectra or at least on spectra at the
termination shock.  Because we do not have spherical geometry which
would result in adiabatic losses (to be addressed in the next phase of
our work), we must artificially impose a free escape boundary to give
the observed high energy cutoff, nevertheless, we feel the most
important difference between our model and previous ones, is that we
treat the injection process in an automatic and more or less
self-consistent fashion.  The efficiency of injection is determined
mostly by the value of the parameter $\eta = \lambda_{\parallel}
/r_{\rm g}$.

To our knowledge, all previous models applied to the termination shock
and based on the transport equation have required particle speeds, $v$,
to satisfy
\begin{equation}
   \lambda_{\parallel}/r_{\rm g} \;\ll\; v/V_{\hbox{\sevenrm sk}}
 \label{eq:isocond}
\end{equation}
or some similar condition. That is, particles that end up as ACRs must
start off with speeds $v \gg V_{\hbox{\sevenrm sk}}$.  This condition
{\it ensures} efficient injection (e.g.  Jokipii 1987), and guarantees
near-isotropy of the distribution functions, a byproduct that permits
use of the diffusion approximation that is central to most transport
equation approaches.  In contrast, our Monte Carlo  technique
effectively finds solutions to the more fundamental Boltzmann equation,
makes no fluid approximations, places no restrictions on the isotropy
of the particle distribution functions, and relates the injection
efficiency to more fundamental aspects of the plasma microphysics.
Moreover, we find that efficient injection is secured in our
simulations, even in nearly perpendicular geometry, when $\eta =
\lambda_{\parallel}/r_{\rm g} \lesssim v/V_{\hbox{\sevenrm sk}}$ is
satisfied, a condition that renders the collision timescale
\teq{\lambda_{\parallel} /v} comparable to or shorter than the time
\teq{r_{\rm g} /V_{\hbox{\sevenrm sk}}} it takes a complete particle
gyro-orbit to convect through the shock.

The automatic nature of injection in our model arises principally
because we assign similar diffusion properties [i.e.
Equation~(\ref{eq:mfp})] to all particles, regardless of whether they are
thermal or highly energetic.   While this differs from other
approaches, we note that for at least some range of particle speeds,
all models of the termination shock must start with an equation similar
to our equation~(\ref{eq:mfp}).  Jokipii \& Giacalone (1996) assume that
\begin{equation}
   \kappa_{\parallel} \; =\; 1.5\times 10^{22} \, \beta \,
   \left ( \dover{R}{10^9 {\rm V}} \right )^{0.5} {\rm cm}^{2}\, {\rm s}^{-1}
\end{equation}
and that $\kappa_{\perp} = 0.1 \kappa_{\parallel}$, where $\beta=v/c$
and $R=pc/(Qe)$ is the particle rigidity in cgs units ($c$ is the speed
of light and  $e$ is the electronic charge).  If the kinetic theory
result $\kappa_{\perp} = \kappa_{\parallel} / (1 + \eta^2)$ is assumed,
this gives $\eta \sim 3$, i.e.  extremely strong cross-field
diffusion.  Chalov \& Fahr (1996a) assume even stronger scattering
(i.e.  $\kappa_{\perp} \simeq \kappa_{\parallel}$ for MeV particles),
while le Roux, Potgieter, \& Ptuskin  (1996) assume
\begin{equation}
   \kappa_{\parallel} = 3.3\times 10^{22}
   \left( \dover{B_1}{10^{-6} \, {\rm G}} \right)^{-1} \! \eta \, \beta
   \left( \dover{R}{10^9 \, {\rm V}} \right) \; {\rm cm}^{2}\, {\rm s}^{-1} ,
 \label{eq:kappaparleroux}
\end{equation}
for $R>0.4$ GV and set $R=0.4$ GV at lower rigidities. Le Roux et al.
also add an extra parameter, $b$, introduced through $ \kappa_{\perp} =
b\, \kappa_{\parallel}/(1 + \eta^2)$ to allow the simultaneous fit to
1987 observations of ACR and galactic cosmic ray spectra and use
$\eta=56$ and $b=47$, giving $\kappa_{\perp} = 0.015
\kappa_{\parallel}$.  This signals a departure from kinetic theory that
presumably might arise with substantial field line wandering.  We
emphasize that in our model, no such added parameters are necessary to
reproduce the ACR hydrogen flux level in the Voyager data.

Through equation~(\ref{eq:mfp}), our model possesses a parallel diffusion
coefficient that is strongly rigidity-dependent for all momenta and is,
in fact, identical to equation~(\ref{eq:kappaparleroux}), including the
numerical coefficient.  Note that contrary to Le Roux et al., we assume
equation~(\ref{eq:kappaparleroux}) holds at {\it all} rigidities.  In any
case, minor differences in the energy dependence and normalization of
$\kappa_{\parallel}$ are unlikely to be important. What we do believe
is important is that, by  including the injection and shock
modification coherently with the acceleration to the highest ACR
energies, we can determine the {\it absolute acceleration efficiency}
as a function of $\eta$ and other parameters. This allows us to
estimate the $\eta$ needed to produce observed ACR intensities and to
relate this microphysical parameter to macrophysical ones (e.g.
$\Theta_{\rm Bn}$ and Mach number).

Our fundamental result is that standard diffusive  shock acceleration
allows  for the injection and acceleration of pickup ions to ACRs
energies with the observed spectral shapes and absolute intensities if
scattering of the strength that is typically assumed in current models
is applied to all particles.  There is no threshold energy or speed
required for shock acceleration to occur. The injection process is a
continuous one with the efficiency being a smoothly increasing function
of the scattering intensity and does not depend critically on any of
the parameters we use. We see no need to invoke complications such as
field line wandering even though it's obvious that if large scale
motions of the magnetic field are present, they may produce modest
changes in the efficiency and modulation (e.g.  le Roux, Potgieter, \&
Ptuskin  1996).  It also seems likely that whatever field line
wandering is present is not self-generated but comes from an
independent background. If the termination shock is producing
self-generated turbulence of the intensities assumed by current models,
this turbulence should be much more intense than any background
turbulence.

We also showed from efficiency considerations that less oblique regions
of the termination shock are less likely to contribute a significant
fraction of the ACRs. Unless much more complicated models are imagined,
the only way to obtain intensities consistent with the observed ACR
intensities and estimates of pickup ion densities at portions of the
termination shock that have $\Theta_{\rm Bn}$ significantly smaller
than $90^\circ$, is by reducing the scattering efficiency, i.e.  by
increasing $\eta$.  Increasing $\eta$ causes time and length scales to
increase and these can become inconsistent with the termination shock
size and charge-stripping rates. This seems to conflict with the
analytic results of Chalov \& Fahr (1996b) and those stemming from one-
or two-dimensional hybrid simulations, which conclude that only regions
with moderate obliquity ($\Theta_{\rm Bn} \lesssim 75^\circ$ for Chalov
\& Fahr and $\Theta_{\rm Bn} \lesssim 60^\circ$ for hybrid results) can
be producing ACRs.

Finally, while we obtain He$^{+}$/H$^{+}$\ and O$^{+}$/H$^{+}$\ ratios
which are somewhat smaller than reported by Cummings \& Stone  (1996),
we definitely see an $A/Q$ enhancement effect during acceleration, as
illustrated in Figure~\ref{fig:SpecFlat}.  We note that the lower Mach number
example tends to produce larger $A/Q$ enhancements due to a complex
interplay between shock parameters such as \teq{\eta}, \teq{\Theta_{\rm
Bn1}}, and Mach number, which we adjust to obtain the same ACR
H$^{+}$\ flux.  For low Mach number shocks, with lower compression
ratios, $\eta$ must be smaller (i.e.  the scattering must be stronger)
to allow increased injection to end up with the observed ACR fluxes.
For our Model III, we obtain enhancements for He$^{+}$\ and
O$^{+}$\ over H$^{+}$\ of $\sim 4$ and 6, respectively, which is about
a factor of 4 less than that inferred by Cummings \& Stone  (1996).  It
is not clear why our enhancements are less, but it may suggest that
there are effects not included in our model which will increase the
acceleration efficiency of heavy ions relative to protons.  However,
adding a cross-shock potential (an obvious extension of our model)
might well lead to enhanced proton acceleration relative to heavier
ions, worsening the discrepancy.  For now, we leave this as an
important unsolved problem.

\section{Conclusions}

In this paper, we have shown that diffusive shock acceleration
operating at the termination shock can account for observed ACR proton
fluxes by directly accelerating pickup ions from solar wind speeds to
$\sim 150$ MeV.  The only requirements for direct injection is that
local magnetic turbulence exists (presumably self-generated) such that
$\kappa_{\perp}/\kappa_{\parallel} \gtrsim 0.01$ and that
$\lambda_{\parallel}$, for pickup ions injected at the shock, is a
small fraction of an AU.  These criteria are not difficult to satisfy
in heliospheric environments, so we suggest that previous work claiming
that a pre-acceleration stage is {\it required} for diffusive shock
acceleration to explain ACR production at the termination shock was
based on incomplete modeling of the acceleration process.  We believe
this is the first calculation of the {\it absolute} intensities of ACRs
using standard solar wind quantities and basic microphysical
parameters.  We find that the acceleration process at the termination
shock is, as far as limited observations allow us to determine,
identical in all important respects to diffusive particle acceleration
observed at inner heliospheric systems such as the Earth bow shock and
travelling interplanetary shocks.

\acknowledgments

We thank Alan Cummings for furnishing Voyager data and Phil Isenberg
and Keith Ogilvie for discussions on pick-up ions.  MGB acknowledges
the support of a Compton Fellowship during the period when most of
this work was completed.  This work was supported by the NASA Space
Physics Theory Program.


\end{document}